\documentclass[12pt]{iopart}
\usepackage{iopams}
\usepackage[square,comma,sort&compress,numbers]{natbib}
\usepackage[colorlinks,linkcolor=blue,anchorcolor=blue,citecolor=blue,filecolor=blue,menucolor=blue,urlcolor=blue]{hyperref}

\newcommand{\steady}{\rho_{eq}}

\newcommand{\da}{\Delta_{\! A}}
\newcommand{\dr}{\Delta_{\rho}}
\newcommand{\dda}{\delta \! A}
\newcommand{\ddr}{\delta \rho}
\newcommand{\mic}{\mathrm{mic}}

\begin{document}

\title{Eigenstate thermalization: Deutsch's approach and beyond}

\author{Peter Reimann}
\address{Universit\"at Bielefeld, Fakult\"at f\"ur Physik, 
33615 Bielefeld, Germany}

\begin{abstract}
The eigenstate thermalization hypothesis (ETH) postulates 
that the energy eigenstates of an isolated many-body
system are thermal, i.e., each of them already yields 
practically the same expectation values as the 
microcanonical ensemble at the same energy.
Here, we review, compare, and extend some recent
approaches to corroborate this hypothesis
and discuss the implications for the system's
equilibration and thermalization.
\end{abstract}

\tableofcontents
\section{Introduction}
\label{s1}
The relaxation of a macroscopic
many-body system towards thermal equilibrium is a very 
common phenomenon, but has still not been satisfactorily 
understood theoretically \cite{skl93,tas98}.
In particular, isolated systems and their text-book 
description at equilibrium by a microcanonical 
ensemble \cite{diu}
have recently regained considerable attention 
\cite{spec,gol10a,pol11}.
An immediate first puzzle is the mere 
fact that the system apparently 
approaches a steady long-time limit
though the quantum mechanical time 
evolution of a non-equilibrium initial 
state is well-known {\em not} to become
asymptotically time-independent 
(see e.g. Sect. \ref{s2} below).
As one possible way out, one could, 
for instance, try to show that after a 
sufficiently long ``equilibration time'', 
the expectation values of pertinent
observables become ``practically constant''
(fluctuations remain below any reasonable 
resolution limit) for ``practically all'' 
later times (exceptions do exists -- 
e.g. due to quantum revivals -- 
but are exceedingly rare).
Indeed, results of this type have 
been established under fairly weak 
and plausible assumptions about 
the initial state, the Hamiltonian, 
and the observables of the considered 
system \cite{tas98,rei08,lin09,sho11,sho12,rei12a,rei12b}.
As an natural next step, quantitative estimates 
of the above mentioned equilibration times 
are currently attracting increasing 
interest \cite{sho12,gol13,mon13,mal14,gol15}.
This is a very important but also quite 
difficult issue of its own right, which goes 
beyond the scope of our present paper.

Here, we rather will focus on another natural 
next issue, named thermalization:
Given that the expectation value of an observable 
equilibrates in the above sense, how 
well does this long-time limit agree with 
the corresponding microcanonical expectation 
value, as predicted by equilibrium statistical 
mechanics?
A sufficient condition for a good such agreement
is the so-called eigenstate thermalization hypothesis 
(ETH), essentially postulating that the expectation 
values of pertinent observables
exhibit negligible variations for all 
energy eigenstates 
with sufficiently close energy eigenvalues.
This hypothesis has its roots in closely 
related conjectures by Berry and Voros
about the energy eigenstates of (fully)
chaotic systems in the semiclassical limit, 
see e.g. Eq. (9) in \cite{ber77} 
and Eq. (6.17) in \cite{vor77}.
Their implications for the 
(diagonal as well as off-diagonal) 
matrix elements 
in energy representation for observables 
with a well-behaved classical limit were
further explored by Feingold and coauthors 
\cite{fei85,fei86,fei89}.
The key role of ETH for thermalization 
in high dimensional chaotic systems 
in the semiclassical regime was first 
recognised by 
Srednicki\footnote{It may be worth noting 
that in those semiclassical studies 
\cite{ber77,vor77,fei85,fei86,fei89,sre94,sre96,sre99}
the term ``microcanonical ensemble'' is used quite
differently than in the present paper (see Sect. \ref{s3}), 
namely referring to classical phase space averages over 
an {\em infinitely thin} energy surface.}
\cite{sre94,sre96,sre99}.
Even earlier, its actual validity was 
numerically (and implicitly) exemplified 
and adopted as an explanation for 
the observed thermalization 
in a spin-chain model by Jensen 
and Shankar 
\cite{jen85}.
More recently, the seminal paper by Rigol, Dunjko, and Olshanii
\cite{rig08} introduced the term ETH, pinpointed its 
importance for thermalization, and stimulated 
numerous, predominantly numerical studies
on the validity of ETH for a large 
variety of specific models 
(mostly spin-chain- or Hubbard-like), 
initial conditions (often involving some quantum quench), 
and observables (mainly few-body or local), 
see e.g. \cite{rig09a,rig09b,san10a,san10b,pal10,rou10,bri10,gog11,ike11,rig12,neu12,ike13,beu14,ste14}.

Mathematically, the validity of ETH could 
be demonstrated so far only in special 
cases, namely for the eigenfunctions of the 
Laplace operator on an arbitrary dimensional
compact Riemannian manifold whose geodesic 
flow is ergodic.
If also the considered observables are 
sufficiently well-behaving, then ETH can be 
proven to hold for the vast majority of all 
eigenfunctions with asymptotically large eigenvalues  
\cite{shn74,col85,zel87}.

Another analytical key work is due to Deutsch,
implicitly verifying ETH for the vast
majority of systems, whose Hamiltonians have 
been sampled according to a certain random matrix 
ensemble
\cite{deu91a,deu91b,deu10}.
Here, we generalise this approach by Deutsch
and unravel its close connection with other recent
explorations of thermalization, especially
by Goldstein and coworkers
\cite{gol13,gol15,gol10a,gol10b,gol11,neu29}.

\section{Equilibration}
\label{s2}
We consider a large (macroscopic but finite), 
isolated system, modelled in terms of a 
Hamiltonian $H$ with eigenvalues $E_n$
and eigenvectors $|n\rangle$, where $n\in{\mathbb{N}}$
and $E_{n+1}\geq E_n$.
System states -- either pure or mixed --
are described by density operators $\rho (t)$, 
evolving according to $\rho(t)=U_t\rho(0) U_t^\dagger$
with propagator $U_t:=\exp\{-\rmi Ht\}$ and $\hbar =1$.
It follows that $\rho_{mn}(t):=\langle m|\rho(t)|n\rangle$
is given by $\rho_{mn}(0) \exp[-\rmi (E_m-E_n)t]$,
i.e., unless the system was already in a steady
state initially, it remains time-dependent
forever. In other words, non-equilibrium 
initial states do not seem to ``equilibrate''
towards a steady long-time limit in an 
obvious way.

Observables are represented by self-adjoint
operators $A$ with expectation values
$\Tr\{\rho(t)A\}$.
In order to model real experimental 
measurements it is, however,
not necessary to admit any arbitrary self-adjoint 
operator \cite{khi60,realobs2,realobs3,realobs4,realobs5,lof,geo95,pop06}.
Rather, it is sufficient to focus on
experimentally realistic observables 
in the following sense 
\cite{rei08,rei08a}: 
Any observable $A$ must represent an experimental 
device with a finite range of possible 
outcomes of a measurement,
\begin{equation}
\da := a_{max}-a_{min} \ ,
\label{a8a}
\end{equation}
where $a_{max}$ and $a_{min}$
are the largest and smallest 
eigenvalues of $A$.
Moreover, this working range $\da$
of the device must be limited to experimentally 
reasonable values compared to its 
resolution limit $\dda$.
Indeed, real measurements usually yield
at most 10-20 relevant digits, i.e.
it is sufficient to consider
range-to-resolution ratios
$\da/\dda \leq 10^{20}$.

Next we define for any given $\dda>0$ and $T>0$
the quantity
\begin{equation}
T_{\dda} := 
\big| \{0 \leq t  \leq T\, : \, |\Tr\{\rho(t)A\}-\Tr\{\steady A\}|\geq\dda\}\big| \ ,
\label{69}
\end{equation}
where $|\{...\}|$ denotes the size (Lebesgue measure) of the 
set $\{...\}$ and where the time-independent,
so-called equilibrium or diagonal ensemble 
$\rho_{eq}$ is defined as the diagonal part 
of $\rho(0)$, i.e. 
$(\rho_{eq})_{mn}:=\delta_{mn}\rho_{nn}(0)$.
As detailed e.g. in \cite{rei12a,rei12b},
one then can show that
\begin{equation}
\frac{T_{\dda}}{T}\leq  2\,\left(\frac{\da}{\dda}\right)^2\, 
{\max_n}\{\rho_{nn}(0)\}
\label{71}
\end{equation}
for all sufficiently large $T$.
For the sake of simplicity, we also
have taken here for granted 
that the energy gaps $E_m-E_n$ are
finite and mutually different for all pairs 
$m\not = n$.
Generalisations have been worked out e.g. in 
\cite{sho11,rei12a,rei12b,sho12}.

According to (\ref{69}),
the left hand side of (\ref{71})
represents the fraction of all times
$t\in[0, T]$, for which there is an experimentally
resolvable difference between the true expectation 
value $\Tr\{\rho(t)A\}$ and the time-independent 
equilibrium expectation value $\Tr\{\steady A\}$.
On the right hand side, $\da/\dda$ is 
the above mentioned range-to-resolution ratio
and ${\max_n} \{\rho_{nn}(0)\}$ 
represents the largest occupation 
probability of an energy eigenstate
(note that the $\rho_{nn}(t)$ are 
conserved quantities).

For a macroscopic $N$-body system
there are roughly $10^{\Or(N)}$ energy 
eigenstates with eigenvalues in every 
interval of $1$J beyond the ground 
state energy \cite{diu}.
Since $N=\Or(10^{23})$, the energy 
levels are thus unimaginably dense 
and even the most careful experimentalist 
will not be able to populate only a few of them
with significant probabilities $\rho_{nn}(0)$.
In the generic case we thus expect
\cite{rei08,rei08a}
that -- even if the system's energy is fixed up 
to an extremely small experimental uncertainty,
and even if the energy levels are populated 
extremely unequally --
the largest population $\rho_{nn}(0)$ 
will still be extremely small 
(compared to $\sum_n \rho_{nn}(0)=1$),
overwhelming by far the factor
$(\da/\dda)^2$ on right hand 
side of (\ref{71}).

Since the level populations $\rho_{nn}(0)$ 
are the result of the system preparation, 
a more detailed understanding and 
quantification of those
terms necessarily requires the modelling of
such a preparation procedure.
We come back to this point in Sect. \ref{s7},
where arguments will be provided that
\begin{equation}
\max_n\{\rho_{nn}(0)\}=10^{-\Or(N)} 
\label{12}
\end{equation}
can be expected in many cases.

From (\ref{71}) together with
(\ref{12}) we can conclude that the system generically
equilibrates in the sense that it
behaves in every possible experimental
measurement exactly as if it were in the 
equilibrium state $\steady$ for the 
overwhelming majority of times within 
any sufficiently large time interval $[0, T]$.

\section{Thermalization}
\label{s3}
Next we address the question whether, and to 
what extent, the above discussed equilibrium 
expectation value $\Tr\{\steady A\}$
is in agreement with the 
corresponding microcanonical expectation 
value, as predicted by the textbooks on 
equilibrium statistical mechanics 
for our isolated $N$-body system at 
hand \cite{diu}.

To begin with, $I_{\mic}:=[E -\Delta E, \,E]$
denotes the usual microcanonical energy
window about the (approximately known 
and thus preset) system energy $E$,
whose width $\Delta E$ is macroscopically
small (below the experimental resolution limit)
but microscopically large (much larger than the
typical energy level spacing $E_{n+1}-E_n$).
The number of energy eigenvalues $E_n$
contained in $I_{\mic}$ is denoted as $D$ 
and is typically very large.
The corresponding microcanonical ensemble 
is given by 
\begin{equation}
\rho_{\mic}:= \frac{1}{D}\sum_{\mic} | n\rangle\langle n| \ ,
\label{z1}
\end{equation}
where the sum $\sum_{\mic}$ runs over all 
$n$ with $E_n\in I_{\mic}$.
In other words, $\rho^{\mic}_{nn}=1/D$
if $E_n\in I_{\mic}$ and 
$\rho^{\mic}_{nn}=0$ otherwise.
Hence, the expectation value of $A$
in the microcanonical ensemble 
takes the form
\begin{equation}
\Tr\{\rho_{\mic} A\} 
=\sum_n \rho^{\mic}_{nn} A_{nn}
=\frac{1}{D}\sum_{\mic} A_{nn} \ .
\label{102}
\end{equation}

On the other hand, recalling that
$(\rho_{eq})_{mn}:=\delta_{mn}\rho_{nn}(0)$
implies
\begin{equation}
\Tr\{\steady A\} =\sum_n \rho_{nn}(0) \, A_{nn} \ .
\label{101}
\end{equation}
As usual, we henceforth assume that the system is
experimentally prepared at the preset macroscopic 
energy $E$,
i.e. also the $\rho_{nn}(0)$'s are negligibly small
for energies $E_n$ outside $I_{\mic}$.
However, within $I_{\mic}$ the actual populations
$\rho_{nn}(0)$ are still largely unknown 
and cannot be controlled by the experimentalist.
In general we therefore have to admit the
possibility that they considerably
vary in a largely unknown (pseudo-random) 
fashion even between neighbouring $n$'s.

The problem of thermalization thus amounts
to showing that the difference between
(\ref{102}) and (\ref{101}) is negligible
in spite of the lack of knowledge about the
$\rho_{nn}(0)$'s.

\section{Eigenstate thermalization hypothesis (ETH)}
\label{s4}
As mentioned in the introduction,
the ETH consists in the surmise
that the expectation values 
$A_{nn}$ of an observable $A$ hardly 
differ for eigenstates $| n\rangle$ 
of a many-body Hamiltonian $H$ with
sufficiently close energy eigenvalues $E_n$ 
\cite{deu91a,deu91b,sre94,sre96,sre99,rig08}.
In particular, if the variations of the $A_{nn}$'s
are negligible over the entire microcanonical 
energy window $I_{\mic}$, then the
(approximate) equality of (\ref{102}) and
(\ref{101}) follows immediately.
In this sense, ETH is a sufficient
(but not necessary) condition for thermalization.

Similarly as for the microcanonical ensemble 
in (\ref{102}), ETH also implies the equivalence 
of $\rho_{eq}$ in (\ref{101}) with a large variety 
of other pure or mixed steady states, 
whose level populations are mainly concentrated 
within the energy window $I_{\mic}$.
On the one hand, this includes other equilibrium
ensembles such as the canonical ensemble,
provided the considered energy interval 
$\Delta E$ is large enough to accommodate 
all notably populated energy levels.
(As we will see later, the latter requirement 
is in fact quite problematic.)
On the other hand, even a single 
energy eigenstate 
$|n\rangle$ with $E_n\in I_{\mic}$
will do.
In other words \cite{rig08}, 
such an energy 
eigenstate encapsulates all 
properties of the considered 
many-body system at 
thermal equilibrium!

Two rather delicate problems, 
which any ``validation'' of ETH 
has to resolve, are as follows:
(i) For any given Hamiltonian $H$, 
one can readily construct (a posteriori)
observables $A$ which violate ETH,
e.g. $A_{nn}=(-1)^n$ and 
arbitrary $A_{mn}$ for $m\not=n$.
(In contrast to what Ref. \cite{beu14}
might suggest, an ETH-violating observable
thus needs {\em not} be a conserved 
quantity.)
In particular, this example implies that
ETH cannot be satisfied simultaneously 
for all observables, and in fact not even 
for all experimentally realistic observables
as specified below Eq. (\ref{a8a}).
(ii) While ETH claims that expectation 
values $A_{nn}$ are (practically) equal
for sufficiently close energy eigenvalues $E_n$,
generically there are -- of course --
notable differences $A_{mm}-A_{nn}$ 
when $E_m-E_n$ is {\em not} small.
But how can the observable $A$ ``feel''
whether the two eigenstates $|m\rangle$ 
and $|n\rangle$ of $H$ belong to similar
energies or not, without any a priori
knowledge about the Hamiltonian $H$ ?

At first glance, it thus might seem 
unavoidable to somehow restrict 
the set of admissible observables.
Indeed, the early explorations of ETH
\cite{ber77,vor77,fei85,fei86,fei89,sre94,sre96,sre99}
solely had in mind semiclassical (small $\hbar$) 
systems, which are classically chaotic,
in conjunction with observables, 
which are $\hbar$-independent and
derive from smooth classical phase 
space functions (see Sect. \ref{s1}).
In contrast, the more recent, predominantly 
numerical studies were mainly focused on 
spin-chain- and Hubbard-like models
\cite{rig09a,rig09b,san10a,san10b,pal10,rou10,bri10,gog11,ike11,ike13,rig12,neu12,beu14,ste14}
(i.e. without an obvious classical limit),
and on few-body or local observables.
Yet another option would be to 
only admit macroscopic observables, 
see Sect. \ref{s8} below.
In either case, it is still not obvious
whether and why such a restricted class 
of observables may get around the above
mentioned problems (i) and (ii).
The solution of those problems 
within our present approach will be 
discussed in Sect. \ref{s9}.

\section{The approach by Deutsch}
\label{s5}
In this section, we reconsider the approach 
by Deutsch, originally published in \cite{deu91a}.
For the detailed calculations, 
announced as Ref. [6] therein, 
see \cite{deu91b}. 
For an updated summary, see 
also \cite{deu10}.

\subsection{Random matrix model}
\label{s5a}
Following Deutsch \cite{deu91a}
we consider Hamiltonians $H$ of the form
\begin{equation}
H=H_0+V \ ,
\label{a10}
\end{equation}
consisting of 
an ``unperturbed'' part $H_0$
and a ``perturbation'' $V$.
As before, eigenvectors and eigenvalues of $H$ are denoted
as $|n\rangle$ and $E_n$ with $E_{n+1}\geq E_n$.
Likewise, those of $H_0$ are denoted as
$|n\rangle_{\! 0}$ and 
$E_n^0$ with $E^0_{n+1}\geq E^0_n$.

Typical examples one has in mind \cite{deu91a}
are $H_0$ which describe a non-interacting
many-body system, e.g. an ideal gas in a box,
while $V$ accounts for the particle-particle 
interactions.
Further examples are so-called quantum quenches:
$H_0$ describes the system for times $t<0$,
while $H$ applies to $t\geq 0$.
In other words, some external condition 
or some system property suddenly changes
at time $t=0$.

In various such examples, the perturbation matrix 
\begin{equation}
V^0_{mn}:=_{\,\, 0\!\!\!}\langle m| V | n\rangle_{\! 0}
\label{a15}
\end{equation}
is often expected or numerically found to 
be a banded matrix
\cite{deu91a,fei89,fyo96,gen12}, 
i.e., the typical magnitude of $V^0_{mn}$
decreases with increasing $|m-n|$ towards zero.
Furthermore, in the above mentioned example
where $H_0$ describes a non-interacting
many-body system, the perturbation matrix 
$V^0_{mn}$ is usually very sparse, i.e.,
only a small fraction of all matrix elements
is non-zero \cite{fyo96,fla97,bro81}.

In any case, the perturbation $V$ is required
to be sufficiently weak so that the two 
systems $H$ and $H_0$ still exhibit 
similar thermodynamic properties at 
the considered system energy $E$,
in particular similar densities of the
energy levels, see above Eq. (\ref{12}).

As a next step, the common lore of random 
matrix theory is adopted \cite{deu91a,gol10b,bro81}: 
One samples matrices $V^0_{mn}$ from
a certain random matrix ensemble with statistical
properties which imitate reasonably well the main 
features of the ``true'' perturbation $V$
(band structure, sparsity etc.), and it is
assumed that if a certain property can be shown
to apply to the overwhelming majority of such 
randomly sampled $V$-matrices, then it will also 
apply to the actual (non-random) $V$ in (\ref{a10}).
A priori, such a random matrix approach may 
appear ``unreasonable'' since most of those 
randomly sampled perturbations $V$ amount
to systems which are physically very different
from the one actually modelled in (\ref{a10}).
Yet, in practice such a random matrix approach 
turned out to be surprisingly successful in a 
large variety of specific examples \cite{bro81}, 
and hence, as in Deutsch's work \cite{deu91a}, 
will be tacitly taken for granted from now on.

\subsection{General framework}
\label{s5b}
The randomness of $V$ entails via in (\ref{a10})
a randomisation of the eigenstates $|n\rangle$ of $H$ 
and hence of the basis-transformation matrix
\begin{equation}
U_{mn}:=\langle m|n\rangle_{\! 0} \ .
\label{a20}
\end{equation}
Likewise, any given observable $A$ and 
its matrix elements in the unperturbed basis
\begin{equation}
A^0_{mn}:=_{\, 0\!\!\!}\langle m| A | n\rangle_{\! 0}
\label{a21}
\end{equation}
are non-random quantities, while 
\begin{equation}
A_{mn}:=\langle m| A | n\rangle
\label{a22}
\end{equation}
will be the elements of a random matrix, 
inheriting the randomness of the $U$-matrix via
\begin{equation}
A_{mn}=\sum_{jk} U_{mj} A_{jk}^0 U_{nk}^\ast \ .
\label{a30}
\end{equation}

Demonstrating ETH thus amounts to
showing that $A_{mm}-A_{nn}$ is small 
for most $V$'s and sufficiently close 
$m$ and $n$. Formally, this will be achieved
by considering the variances
\begin{equation}
\sigma^2_n :=
\langle (A_{nn} - \langle A_{nn}\rangle_V)^2\rangle_V=
\langle (A_{nn})^2\rangle_V - \langle A_{nn}\rangle_V^2 \ ,
\label{a50}
\end{equation}
where $\langle ... \rangle_V$ indicates the 
average over the random perturbations $V$.
In a first step (Sect. \ref{s5d}), 
we will show that the mean values 
$\langle A_{nn}\rangle_V$ 
for sufficiently close $n$'s
differ very little
in comparison to the experimental resolution 
limit $\dda$ 
introduced in Sect. \ref{s2}.
In a second step (Sect. \ref{s5e}), 
we will show that
$\sigma_n \ll \dda$, implying that $A_{nn}$
differs very little from $\langle A_{nn}\rangle_V$ for
most $V$.
Altogether, this will imply the desired result
that for most $V$'s the $A_{nn}$'s change
very little upon changing $n$ (Sect. \ref{s5f}).

To simplify the algebra, we henceforth
assume that the largest and smallest
eigenvalues of $A$ in (\ref{a8a}) 
satisfy 
\begin{equation}
a_{min}=-a_{max} \ .
\label{a51}
\end{equation}
As a consequence, $a_{max}=\da/2$
according to (\ref{a8a}).
Note that the assumption (\ref{a51}) 
does not imply any loss of generality, 
since adding an arbitrary constant 
(times the identity operator) to 
the observable $A$,
and hence to all its eigenvalues,
does not entail any non-trivial 
physical consequences.
In particular, the above mentioned 
changes of $\langle A_{nn}\rangle_V$ 
upon variation of $n$ and the variances
(\ref{a50}) remain exactly the same.
For later use, we thus can conclude 
that
\begin{equation}
|\langle\psi |A^\nu| \psi\rangle|\leq (\da/2)^\nu
\label{a8}
\end{equation}
for any $\nu\in\mathbb{N}$.

For the sake of simplicity, we furthermore
assume that all matrix elements $V^0_{mn}$ 
from (\ref{a15}) and $U_{mn}$ from (\ref{a20}) 
are real numbers. 
For example, for systems without
spins and magnetic fields,
$H_0$ and $V$ in (\ref{a10}) 
are both purely real operators in position 
representation and hence the eigenstates
$|n\rangle_0$ and $|n\rangle$ 
can be chosen so that all $V^0_{mn}$ 
and $U_{mn}$ become real.
So, it is natural to assume that also
the corresponding random matrix ensembles
only involve real matrix elements.
In particular, this implies with Eqs.
(\ref{a30}) and (\ref{a50}) that
\begin{eqnarray}
\langle A_{nn}\rangle_V & = & \sum_{jk} A_{jk}^0 \langle U_{nj} U_{nk}\rangle_V
\label{a60}
\\
\sigma^2_n & = & \sum_{ijkl} A_{ij}^0 A_{kl}^0 
\langle U_{ni} U_{nj} U_{nk} U_{nl} \rangle_V 
- \langle A_{nn}\rangle_V^2 \ .
\label{a70}
\end{eqnarray}

It may well be that our subsequent calculations
can be readily extended to systems for which 
such a transformation to purely real matrices 
$V^0_{mn}$ and $U_{mn}$ is no longer possible.
However, the so far available knowledge,
e.g. from random matrix theory or numerical investigations,
regarding the statistical properties of the $U_{mn}$'s 
is only sufficient for our purposes for real matrices
(see next Section).
Since the subject of our paper is not the
exploration of such statistical random matrix
properties but rather their implications with 
respect to ETH, we confine ourselves to the 
case of real matrices.

\subsection{Properties of $U_{mn}$}
\label{s5c}
In view of (\ref{a60}), (\ref{a70}), 
some basic statistical properties of 
the matrix elements $U_{mn}$ are 
needed in order to
make any further progress.

At this point it is crucial to note
that the Hamiltonian $H$ in (\ref{a10}) 
gives rise to a very
special type of random matrix.
Namely, the matrix 
$_{\,\, 0\!}\langle m| H | n\rangle_{\! 0}$
is the sum of the above discussed 
random perturbation $V^0_{mn}$
and of the non-random diagonal matrix
$_{\,\, 0\!}\langle m| H_0 | n\rangle_{\! 0}=\delta_{mn}E_{n}^0$,
whose diagonal elements $E_n^0$ 
grow approximately linearly with $n$
(at least within a sufficiently small vicinity
of the preset system energy $E$,
onto which we tacitly restrict ourselves,
see also Sects. \ref{s3} and \ref{s5f}).
Out of the huge literature on random matrix
theory, only a relatively small number 
of works pertains to this special case,
see e.g. \cite{deu10,fyo96,fla97,cas96} 
and further references therein.
Strictly speaking, they are obtained for infinitely large
matrices $V^0_{mn}$, whose statistical properties
do not depend on $m$ and $n$ separately, but only
on the difference $m-n$. 
Likewise, the unperturbed matrix 
$_{\,\, 0\!}\langle m| H_0 |n\rangle_{\! 0}$
is assumed to be infinitely large and of the form $\delta_{mn}E^0_n$
with equally spaced energy gaps $E^0_{n+1}-E^0_n$.
Intuitively, these seem quite plausible approximations,
at least for not too strong perturbations $V$ in (\ref{a10}).
They can be readily justified by numerical examples,
but somewhat more rigorous analytical results do not 
seem to exist.
Here, we adopt the widely accepted viewpoint that,
for out present purposes,
they can be taken for granted \cite{deu91a,fyo96,bro81}.

As a consequence, also the statistical properties of the
$U_{mn}$ only depend on $m-n$, e.g. the $\nu$-th moments 
are of the form 
\begin{equation}
\langle(U_{mn})^\nu\rangle_V=u_\nu(m-n) \ ,
\label{a80}
\end{equation}
where the $u_\nu(n)$ are real 
(but not necessarily even \cite{cas96}) 
functions of $n$, and are furthermore
non-negative for even $\nu$.

Known analytical results mainly concern the second 
moment $u_2(n)$ for various ensembles of possibly 
banded and/or sparse random $V$-matrices,
see e.g. \cite{deu10,fyo96,fla97,cas96} 
and references therein.
In all cases, it is found that $u_2(n)$ 
is monotonically decreasing for $n \geq 0$ and
monotonically increasing for $n\leq 0$,
hence exhibiting a global maximum at $n=0$:
\begin{equation}
\max_{n}\{u_2(n)\}=u_2(0) \ .
\label{a101}
\end{equation}
Since $\sum_k U_{mk}U_{nk}=\delta_{mn}$
we can conclude that
\begin{equation}
\sum_n u_2(n)=1 \ ,
\label{a90}
\end{equation}
implying that $u_2(n)$ must approach zero 
for large $|n|$.

In all those analytical results, 
the mean values $\langle V^0_{mn}\rangle_V$
are tacitly assumed to vanish and it is found that 
$u_2(n)$ then only depends on the second 
moments $\langle (V^0_{mn})^2\rangle_V$.
Since we are not aware of any justification for 
this assumption $\langle V^0_{mn}\rangle_V=0$, 
we have numerically investigated various 
examples and found that, indeed,
the statistical properties of the $U_{mn}$'s 
seem to be independent of the first moments
$\langle V^0_{mn}\rangle_V$
(while keeping all other cumulants fixed).
Furthermore, a simple physical argument 
is as follows:
Replacing an eigenstate $|n\rangle_{\! 0}$ 
of $H_0$ by $-|n\rangle_{\! 0}$ is supposed 
not to change any physically relevant properties 
of the given (non-random) model in (\ref{a10}). 
Note that this argument applies 
separately to any single $n$.
Hence, it is quite plausible that upon randomly 
flipping the signs for half of all $n$'s, the
resulting ``new'' $V^0_{mn}$'s will
be ``unbiased'' for $m\not =n$.
(More precisely: if a random matrix 
description works at all, then an ensemble 
with $\langle V^0_{mn}\rangle_V=0$ 
seems most appropriate).
Finally, a possibly remaining systematic ``bias'' 
of the diagonal elements $V^0_{nn}$ can be 
removed by adding an irrelevant constant 
to $V$. (The typical magnitude of the 
$V^0_{mn}$'s is also 
estimated in Appendix B of \cite{deu91b}, 
however, not taking into account the 
possibility that their average may be zero.)

For the rest, the detailed properties of 
$u_2(n)$ are found -- as expected -- to 
still depend on the quantitative details 
of $\langle (V^0_{mn})^2\rangle_V$.
Since no general statement about the 
latter seems possible for the general 
class of systems we have in mind with 
(\ref{a10}), we will focus on conclusions
which do not depend on the corresponding
details of $u_2(n)$.
Rather, we will only exploit the following
very crude common denominator of all so
far explored particular classes of 
random matrices $V_{mn}$,
see e.g. \cite{deu10,fyo96,fla97,cas96} 
and further references therein:
\begin{equation}
u_2(0)=10^{-\Or(N)} \ .
\label{a100}
\end{equation}
The basic physical reason is that exceedingly 
``weak'' perturbations $V$ in (\ref{a10}) 
are tacitly ignored so that the smallest 
relevant energy scale is the mean 
level spacing $E_{n+1}-E_n$, being of 
the order of $10^{-\Or(N)}\,$J 
according to Sect. \ref{s2}.
Moreover, the ratio between this energy scale 
and any other relevant energy scale of the 
system can be very roughly estimated 
by $10^{-\Or(N)}$, independently
of any further details of the specific 
model system in (\ref{a10}).
As a consequence, also the very crude estimate
(\ref{a100}) is independent of these details.

Further statistical properties of the $U_{mn}$,
which we will, similarly as in
\cite{deu91a,deu91b,fyo96,fla97,cas96},
take for granted later on, are:

(i) Their average is zero, i.e.,
\begin{equation}
u_1(n)=0 \ \ \mbox{for all $n$.}
\label{a110}
\end{equation}

(ii) They are statistically independent of each other, i.e.,
\begin{equation}
\mbox{$U_{mn}$ is independent of $U_{jk}$
if $m\not=j$ or $n\not=k$.}
\label{a120}
\end{equation}

(iii) Their distribution does not exhibit long tails, i.e.,
\begin{equation}
u_4(n)\leq c\, u_2(n)\, u_2(0) \ \ \mbox{for all $n$}
\label{a115}
\end{equation}
with an $n$-independent constant $c$, which 
may possibly be very large but which is required
not to be so large that it can compete in order of 
magnitude with $1/u_2(0)$ from (\ref{a100}).
For instance for a system with $N=10^{23}$ particles,
it would be sufficient that $c\leq10^{10^{22}}$.
In other words, we adopt the very weak assumption 
\begin{equation}
c \, u_2(0)=10^{-\Or(N)} \ .
\label{a116}
\end{equation}
E.g., for a Gaussian distribution (with zero mean, 
cf. (\ref{a110})) one finds that 
$u_4(n)=3u_2^2(n)$ and hence 
(\ref{a115}) is satisfied for $c=3$.
Though a Gaussian distribution is often 
taken for granted \cite{deu91a,deu91b,sre94}, 
non-Gaussian distributions have been actually observed 
e.g. in \cite{gen12} and also in our own numerical 
explorations (unpublished),
but the more general condition (\ref{a115})
was still satisfied in all cases.
We also note that since $u_2(n)/u_2(0)$ approaches 
zero for large $|n|$, the condition in 
(\ref{a115}) becomes weaker and weaker with 
increasing $|n|$.

\subsection{Mean values}
\label{s5d}
Evaluating (\ref{a60}) by means of (\ref{a80}), (\ref{a110}), 
and (\ref{a120}) yields
\begin{eqnarray}
\langle A_{nn}\rangle_V & =  & 
\sum_{j} A_{jj}^0 \langle U_{nj} U_{nj}\rangle_V +
\sum_{j\not=k} A_{jk}^0 \langle U_{nj}\rangle_V \langle U_{nk}\rangle_V
\nonumber
\\
& = & 
\sum_{j} A_{jj}^0 u_2(n-j) \ .
\label{a130}
\end{eqnarray}
It follows that
\begin{eqnarray}
\Delta_n
 & := & \langle A_{n+1,n+1}\rangle_V-\langle A_{nn}\rangle_V
\nonumber
\\
& = & \sum_{j} A^0_{jj} [u_2(n+1-j)-u_2(n-j)]
\label{a135}
\end{eqnarray}
and hence that
\begin{eqnarray}
|\Delta_n| & \leq &
\sum_{j} |A^0_{jj}| |u_2(n+1-j)-u_2(n-j)|
\nonumber
\\
& \leq &
\max_j |A^0_{jj}|
\sum_k |u_2(k+1)-u_2(k)| \ .
\label{a139}
\end{eqnarray}
The maximum over $j$ can be estimated from
above by $\da/2$ according to (\ref{a8}).
Recalling that $u_2(k)$ is monotonically 
decreasing for $k\geq 0$ and monotonically
increasing for $k\leq0$ (see above (\ref{a101})),
the sum over $k$ amounts to $2\, u_2(0)$
(more generally, this sum amounts to the total 
variation of $u_2(k)$; hence, if $u_2(k)$ 
exhibits $M$ local maxima, it can be estimated
from above by $2M\max_k u_2(k)$).
Altogether, we thus can conclude that
\begin{equation}
|\langle A_{n+1,n+1}\rangle_V-\langle A_{nn}\rangle_V| 
\leq \da\, u_2(0) \ .
\label{a140}
\end{equation}
This upper bound is tight: One can readily find 
examples $A$ for which (\ref{a140}) becomes an 
equality.
Moreover, it follows that
\begin{equation}
|\langle A_{mm}\rangle_V-\langle A_{nn}\rangle_V| 
\leq |m-n|\, \da \, u_2(0) \ .
\label{a145}
\end{equation}
By generalising the line of reasoning in
(\ref{a139}), (\ref{a140}) one can also 
show that
\begin{equation}
|\langle A_{mm}\rangle_V-\langle A_{nn}\rangle_V|
\leq \da\sum_{k=-\kappa}^\kappa u_2(k)
\label{a146}
\end{equation}
with $\kappa:=|m-n|-1$. 
Under certain conditions, the bound
(\ref{a145}) may be better than (\ref{a146}),
but never by more than a factor of $2$.
For sufficiently large $|m-n|$, (\ref{a146}) is always
better since the sum on the 
right hand side is bounded by unity (see (\ref{a90})),
while the right hand side of (\ref{a145}) is unbounded.
(The relevance of large $|m-n|$-values will 
become apparent in Sect. \ref{s5f} below).
In any case, (\ref{a146}) is a rather tight bound
in the sense that one can find examples for
$A_{jj}^0$ so that the left hand side is larger
than the right hand side divided by $2$ for
a set of suitably chosen pairs $(m,\, n)$
so that the differences $m-n$ may still take 
any integer value.

\subsection{Variances}
\label{s5e}
We rewrite the variance from (\ref{a70}) as
\begin{equation}
\sigma^2_n + \langle A_{nn}\rangle_V^2 
=  \sum_{ijkl} A_{ij}^0 A_{kl}^0 
\langle U_{ni} U_{nj} U_{nk} U_{nl} \rangle_V 
\label{a150}
\end{equation}
and evaluate the four-fold sum by 
distinguishing 4 possible cases.
Case 1: $i\not=k$ and $i=j$. 
In this case, we only have to keep summands
with $l=k$: otherwise the factor $U_{nk}$ on the right hand side of 
(\ref{a150}) would be independent of the remaining three factors
according to (\ref{a120}), and the corresponding summand 
would vanish according to (\ref{a80}) and (\ref{a110}).
Case 2: $i\not=k$ and $i\not=j$. As before, we can conclude that
only summands with $l=i$ and $j=k$ give rise to
non-vanishing terms.
Case 3: $i=k$ and $i\not=l$, implying, as before, that only $j=l$ contribute.
Case 4: $i=k$ and $i=l$, implying $j=i$.
Consequently, we can rewrite (\ref{a150}) as
\begin{equation}
\sigma^2_n + \langle A_{nn}\rangle_V^2   
=  \Sigma_1 + \Sigma_2 + \Sigma_3 + \Sigma_4 \ ,
\label{a160}
\end{equation}
where the four summands correspond to the above four
cases and can be rewritten as:
\begin{eqnarray}
\Sigma_1 & = &  \sum_{i\not=k} A_{ii}^0 A_{kk}^0 
\langle U_{ni} U_{ni} U_{nk} U_{nk} \rangle_V 
\label{a170}
\\
\Sigma_2 & = &  \sum_{i\not=k} A_{ik}^0 A_{ki}^0 
\langle U_{ni} U_{nk} U_{nk} U_{ni} \rangle_V 
\label{a180}
\\
\Sigma_3 & = &  \sum_{i\not=l} A_{il}^0 A_{il}^0 
\langle U_{ni} U_{nl} U_{ni} U_{nl} \rangle_V 
\label{a190}
\\
\Sigma_4 & = &  \sum_{i} A_{ii}^0 A_{ii}^0 
\langle U_{ni} U_{ni} U_{ni} U_{ni} \rangle_V  \ .
\label{a200}
\end{eqnarray}
With the help of (\ref{a120}) and (\ref{a80}) 
we can rewrite (\ref{a170}) as
\begin{eqnarray}
\Sigma_1 & = &
\sum_{i\not=k} A_{ii}^0 A_{kk}^0  u_2(n-i) u_2(n-k)
\nonumber
\\
& = &
\sum_{ik} A_{ii}^0 A_{kk}^0  u_2(n-i) u_2(n-k) - \sum_i (A^0_{ii})^2 u_2^2(n-i)
\nonumber
\\
& = &
\langle A_{nn}\rangle_V^2 - \sum_i (A^0_{ii})^2 u_2^2(n-i) \ ,
\label{a210}
\end{eqnarray}
where we exploited (\ref{a130}) in the last equation.
Likewise, one finds that
\begin{eqnarray}
\Sigma_2 & = & \sum_{ik} |A_{ik}^0|^2 u_2(n-i)u_2(n-k)-\sum_i (A_{ii}^0)^2 u_2^2(n-i)
\label{a220}
\\
\Sigma_3 & = & \sum_{il} (A_{il}^0)^2 u_2(n-i)u_2(n-l)-\sum_i (A_{ii}^0)^2 u_2^2(n-i)
\label{a230}
\\
\Sigma_4 & = & \sum_{i} (A_{ii}^0)^2 u_4(n-i) \ .
\label{a240}
\end{eqnarray}
Introducing these results into (\ref{a160}) thus yields
\begin{eqnarray} 
\sigma^2_n & = & S_1+S_2+S_3
\label{a250}
\\
S_1 & := &   \sum_{ik} |A_{ik}^0|^2 u_2(n-i)u_2(n-k)
\label{a260}
\\
S_2 & := &   \sum_{il} (A_{il}^0)^2 u_2(n-i)u_2(n-l)
\label{a270}
\\
S_3 & := &  \sum_{i} (A_{ii}^0)^2 [u_4(n-i) - 3 u_2^2(n-i)] \ .
\label{a280}
\end{eqnarray}
The three factors on the right hand side of  (\ref{a260}) 
are all non-negative and hence
\begin{equation}
S_1 =  |S_1| 
\leq \max_k\{ u_2(k)\} \sum_i u_2(n-i)\sum_k A^0_{ik}A^0_{ki} \ .
\label{a285}
\end{equation}
With (\ref{a21}) one readily sees that the
last sum over $k$ amounts to 
$_{\,\, 0\!}\langle i | A^2|i\rangle_{\! 0}=:(A^2)^0_{ii}$.
Exploiting (\ref{a101}) we thus obtain
\begin{equation}
S_1 \leq u_2(0) \sum_i (A^2)^0_{ii}\, u_2(n-i) \ .
\label{a290}
\end{equation}
Likewise, since $u_2(i)\geq 0$ for all i,
the modulus of (\ref{a270}) can be estimated as
\begin{equation}
|S_2| \leq  \sum_{i l} |A_{il}^0|^2 u_2(n-i)u_2(n-l) = S_1 \ .
\label{a300}
\end{equation}
Turning to (\ref{a280}), we first note that the last factor
$\kappa_4:= u_4(n-i) - 3 u_2^2(n-i)$ represents the 
4th cumulant of the random variable $U_{ni}$.
For a Gaussian distribution 
(the case considered by Deutsch \cite{deu91a,deu91b}),
this cumulant vanishes, 
but for more general distributions it may be finite
and of either sign. 
We thus estimate $|\kappa_4|$ from above
by $u_4(n-i)+3 u_2^2(n-i)$.
Observing (\ref{a116}) and 
$u_2^2(n-i)\leq u_2(0) u_2(n-i)$
(see (\ref{a101})), we
thus can bound (\ref{a280}) by
\begin{equation}
|S_3| 
\leq u_2(0) (c+3) \sum_{i} (A^0_{ii})^2 u_2(n-i) \ .
\label{a310}
\end{equation}
Next, we invoke the Cauchy-Schwarz inequality
to conclude 
\begin{equation}
\langle\psi | B|\psi\rangle^2
\leq \langle\psi|\psi\rangle
\langle\psi | B^2 |\psi\rangle
\label{a312}
\end{equation}
for arbitrary Hermitian operators $B$ 
and vectors $|\psi\rangle$.
In particular, it follows that 
$(A^0_{ii})^2\leq (A^2)^0_{ii}$.
Altogether, we thus arrive at
\begin{equation}
\sigma^2_n \leq  (c+5)\, u_2(0) \sum_{i} (A^2)^0_{ii} \, u_2(n-i) \ .
\label{a313}
\end{equation}
Finally, we exploit (\ref{a8}) and (\ref{a90}), resulting in
\begin{equation}
\sigma^2_n \leq  (c+5)\, u_2(0) \, (\da/2)^2
= 10^{-\Or(N)}\, \da^2 \ ,
\label{a330}
\end{equation}
where we used (\ref{a116}) in the last step.

\subsection{Discussion}
\label{s5f}
From (\ref{a50}), (\ref{a330}), and
Markov's inequality it follows that
\begin{equation}
\mbox{Prob}
\left(|A_{nn}-\langle A_{nn}\rangle_V|\geq\epsilon\right)
\leq (\da/\epsilon)^2\, 10^{-\Or(N)}
\label{a340}
\end{equation}
for any $\epsilon>0$, where Prob$(X)$ 
denotes the probability that a randomly sampled 
$V$ in (\ref{a10}) entails property $X$.
For instance, if in the last term $\Or(N)=10^{23}$ 
and $\epsilon=\da\, 10^{-10^{22}}$
then the right hand side of (\ref{a340}) is still
$10^{-\Or(10^{23})}$.
Consequently, the joint probability that every $A_{nn}$
is practically indistinguishable from 
$\langle A_{nn}\rangle_V$ simultaneously for 
all $n\in\{n_0,..,n_0+\Delta n\}$ still remains 
negligibly small if $0\leq \Delta n\ll 10^{\Or(N)}$.

On the other hand, (\ref{a100}) and (\ref{a145}) 
imply that the difference 
$\langle A_{mm}\rangle_V-\langle A_{nn}\rangle_V$
remains below the experimental resolution limit 
$\dda$ of $A$ (cf. Sect. \ref{s2})
even for quite large range-to-resolution ratios $\da/\dda$,
provided $|m-n|$ remains much smaller than of 
the order of $10^{\Or(N)}$.
In other words, also the variations of the
$\langle A_{nn}\rangle_V$ remain negligibly 
small within the ``window'' of $n$-values
$\{n_0,..,n_0+\Delta n\}$ if $0\leq \Delta n\ll 10^{\Or(N)}$.

Altogether, we thus arrive at the conclusion that
for the vast majority of randomly 
sampled perturbations $V$ in (\ref{a10}),
the $A_{nn}$'s remain practically
constant (below the experimental resolution limit)
as long as $n$ varies by much less than
$10^{\Or(N)}$.

The latter property is sometimes referred to 
as the {\em strong ETH} \cite{bri10,ike13}.
It immediately implies the practical
indistinguishability of the two expectation
values (\ref{102}) and (\ref{101}), 
and hence thermalization,
provided both the $\rho_{nn}(0)$ and the
$\rho^{\mic}_{nn}$ are negligibly small 
outside a window of $n$-values 
much smaller than $10^{\Or(N)}$,
but otherwise without any further 
restriction on the initial condition 
$\rho(0)$.

If the range $\Delta n$ of admitted $n$-values is not
any more much smaller than $10^{\Or(N)}$,
then we can no longer conclude from (\ref{a330})
that with high probability all the $A_{nn}$'s
remain simultaneously close to the 
$\langle A_{nn}\rangle_V$'s.
However, 
we still can conclude that for the
vast majority of $n$'s, those 
differences remain negligibly 
small\footnote{With $x_n:=|A_{nn}-\langle A_{nn}\rangle_V|$ and
$\delta:=(\da/\epsilon)^2\, 10^{-\Or(N)}$, Eq. (\ref{a340}) can
be rewritten as $\langle\Theta(x_n-\epsilon)\rangle_V\leq\delta$,
where $\Theta(x)$ is the Heaviside step function.
Furthermore, $Z_\epsilon:=\sum_{n=1}^D\Theta(x_n-\epsilon)$ 
counts how many of the $x_n$'s exceed $\epsilon$.
It follows that $\langle Z_\epsilon\rangle_V\leq D\, \delta$ and with
Markov's inequality that $\mbox{Prob}(Z_\epsilon\geq qD)\leq \delta/q$,
where $\mbox{Prob}(Z_\epsilon\geq qD)$ is the probability 
that more than a fraction $q$ of all $x_n$'s exceed $\epsilon$.}.
If, in addition, also the variations of the 
$\langle A_{nn}\rangle_V$'s would remain small,
we still could conclude that ``most''
$A_{nn}$'s are practically equal
(for the overwhelming majority of $V$'s),
i.e. the so-called {\em weak ETH}
is satisfied \cite{bri10,ike13}.
As a consequence, the expectation values
(\ref{102}) and (\ref{101}) would again be
practically equal under certain 
additional conditions on the initial
condition $\rho(0)$.
For instance, the total weight of all 
$\rho_{nn}(0)'s$ corresponding to 
exceptionally large differences
$A_{nn}-\langle A_{nn}\rangle_V$ 
should remain sufficiently small.
E.g. (\ref{12}) would obviously
be a sufficient condition.

However, this line of reasoning contains 
a problem:
If the variations $\Delta n$ of $n$ are 
not any more much smaller 
than $10^{\Or(N)}$, then
Eqs. (\ref{a145}) and (\ref{a100})
no longer imply that the variations
of $\langle A_{nn}\rangle_V$ 
remain negligible.
The same conclusion follows
from the bound (\ref{a146}).
Since the latter bound is already 
rather tight (see below (\ref{a146})),
we can conclude that the restriction
to windows of $n$-values much smaller than 
$10^{\Or(N)}$ is not merely a technical 
problem but rather an indispensable 
prerequisite of the random matrix 
model from Sect. \ref{s5a}.
In particular, this restriction also concerns the
original findings by Deutsch \cite{deu91a}.

Note that $\Delta n$ from above
is identical to the number $D$ of energy
eigenvalues $E_n$ contained in the 
microcanonical energy 
window $I_{\mic}:=[E -\Delta E, \,E]$ from 
Sect. \ref{s3}.
The above discussed restriction thus 
amounts to 
\begin{equation}
D\ll 10^{\Or(N)} 
\label{z100}
\end{equation}
and implies that $\Delta E$ must remain very much
smaller than any macroscopically resolvable energy 
difference. (This follows from the fact that
the energy eigenstates are exponentially 
dense in the system size $N$, see Sect. \ref{s2}).

\section{Srednicki's ETH for the off-diagonal matrix elements}
\label{s6}
In Refs. \cite{sre96,sre99}, Srednicki 
formulated, besides the so far considered 
ETH for the diagonal matrix elements $A_{nn}$,
also a corresponding ETH for the 
off-diagonal elements $A_{mn}$ with $m\not=n$.
This hypothesis can also be readily confirmed 
within our present framework:

Similarly as in (\ref{a60}), we find from (\ref{a30})
that
\begin{equation}
\langle A_{mn}\rangle_V = \sum_{jk} A_{jk}^0 \langle U_{mj} U_{nk}\rangle_V \ .
\label{b10}
\end{equation}
For $m\not=n$,
the last factor $\langle U_{mj} U_{nk}\rangle_V$ equals
$\langle U_{mj}\rangle_V \langle U_{nk}\rangle_V$
according to (\ref{a120}), and hence vanishes 
according to (\ref{a80}) and (\ref{a110}).
In conclusion
\begin{equation}
\langle A_{mn}\rangle_V = 0
\label{b20}
\end{equation}
for all $m\not=n$.

Turning to the second moment 
(variance), one finds similarly as in 
(\ref{a50}), (\ref{a70}) 
that
\begin{eqnarray}
& & \langle |A_{mn}|^2 \rangle_V
= \langle A_{mn} A_{nm} \rangle_V 
=\sum_{ijkl} A_{ij}^0 A_{kl}^0 
\langle U_{mi} U_{nj} U_{nk} U_{ml} \rangle_V \ .
\label{b30}
\end{eqnarray}
According to (\ref{a120}), for $m\not=n$
the last term
$\langle U_{mi} U_{nj} U_{nk} U_{ml} \rangle_V$ 
now factorises into
$\langle U_{mi} U_{ml} \rangle_V
\langle U_{nj} U_{nk}\rangle_V$.
Exploiting (\ref{a120}) once more implies
$\langle U_{mi} U_{ml} \rangle_V=
\delta_{il}\, \langle U^2_{mi}\rangle_V$
and 
$\langle U_{nj} U_{nk}\rangle_V=\delta_{jk}
\langle U^2_{nk}\rangle_V$.
With (\ref{a80}) we thus obtain
\begin{equation}
\langle |A_{mn}|^2 \rangle_V
=
\sum_{ik} |A_{ik}^0|^2 u_2(m-i) u_2(n-k) \ .
\label{b50}
\end{equation}
Step by step as in (\ref{a260}), (\ref{a285}), 
(\ref{a290}), (\ref{a330}) it follows that
\begin{equation}
\langle |A_{mn}|^2 \rangle_V \leq u_2(0) (\da/2)^2
\label{b60}
\end{equation}
for all $m\not=n$.
In view of (\ref{a100}) we see that the
off-diagonals $|A_{mn}|$ are
typically exponentially small
in the system size $N$,
in agreement with Srednicki's 
prediction in Refs. \cite{sre96,sre99}.

The overall conclusion applying to any 
given Hermitian operator $A$ of finite 
range $\da$ is:
The representation of $A$ in the 
eigenbasis of $H$ is, for the overwhelming 
majority of randomly sampled 
perturbations $V$ in (\ref{a10}), 
very close to a diagonal matrix,
whose diagonal elements $A_{nn}$
change very slowly with $n$.

\section{Implications for the level populations $\rho_{nn}(0)$}
\label{s7}
Throughout this section, we consider the density
operator $\rho(0)$ (pure or mixed state)
from Sect. \ref{s2} and abbreviate it as $\rho$.

Since $\rho$ is a Hermitian operator, all results
so far for general observables $A$ are immediately 
applicable to $\rho$. 
However, this particular observable $A=\rho$ 
also exhibits some subtle special features.
Therefore, we first focus on a 
simple example.

\subsection{Simple example}
\label{s7a}
We consider a pure energy eigenstate
of the unperturbed Hamiltonian 
$H_0$ in (\ref{a10}), i.e. 
\begin{equation}
\rho=|m\rangle_{\! 0 \, 0\!}\langle m |
\label{c1}
\end{equation}
with an arbitrary but fixed $m$.
Its eigenvalues are either zero or one, 
hence the range from (\ref{a8a}) is 
$\dr=1$.
Observing that $\rho^0_{ik}=\delta_{im}\delta_{km}$
it follows from (\ref{a30}) and (\ref{a60})-(\ref{a80})
 that
\begin{eqnarray}
\rho_{nn} & = &  |U_{mn}|^2
\label{c2}
\\
\langle \rho_{nn}\rangle_V & = & u_2(n-m)
\label{c4}
\\
\sigma^2_n & = & u_4(n-m)-u^2_2(n-m) \ ,
\label{c6}
\end{eqnarray}
where the variance from (\ref{a50}) is given for $A=\rho$ by 
\begin{equation}
\sigma^2_n := 
\langle(\rho_{nn})^2\rangle_V - \langle \rho_{nn}\rangle_V^2 \ .
\label{c8}
\end{equation}
As a concrete example, we may focus on Gaussian 
distributed 
$U_{mn}$'s (see below Eq. (\ref{a116})), so that
$u_4(n)=3 u_2^2(n)$ and hence
\begin{equation}
\sigma^2_n = 2\, u^2_2(n-m) \ .
\label{c10}
\end{equation}

Altogether, the standard deviation $\sigma_n$
of the random variable $\rho_{nn}$ from
(\ref{c2}) is thus comparable to
its mean value $\langle\rho_{nn}\rangle_V$, 
and both are,
according to (\ref{a101}) and (\ref{a100}),
extremely small compared to the range $\dr=1$
of the considered observable $\rho$.
Within any reasonable resolution limit $\ddr$
of this observable we thus can conclude
that, for the vast majority of random 
perturbations $V$ in (\ref{a10}), 
all $\rho_{nn}$'s are practically
equal (namely zero),
in agreement with the general validation of 
ETH from Sect. \ref{s5}.
But for our present purposes, this usual
resolution limit $\ddr$ is still way too large.
On the actual scale of interest, the $\rho_{nn}$'s 
from (\ref{c2}) are not at all a slowly 
varying function of $n$, but rather
exhibit very significant 
random fluctuations (see also (\ref{a120})).
In particular, it would be wrong to argue 
that the $\rho_{nn}(0)$'s in (\ref{101})
are now practically constant and
hence, upon comparison 
with (\ref{102}), thermalization follows.

\subsection{General case}
\label{s7b}
We return to general density operators 
$\rho$, i.e., we only assume that $\rho$ is
a Hermitian, non-negative operator
of unit trace and purity
\begin{equation}
\Tr\{\rho^2\} \leq  1 \ .
\label{c12}
\end{equation}
In the following, we will exploit these 
properties of $\rho$, which, however,
would be lost after adding a constant 
to $\rho$ so that (\ref{a51}) is satisfied.
Hence we only can employ those
previous results which were obtained
without the help of (\ref{a51}).
Along these lines, one finds exactly 
as in (\ref{a130}) and (\ref{a313}) 
that
\begin{eqnarray}
\langle \rho_{nn}\rangle_V & =  &
\sum_{j} \rho_{jj}^0 u_2(n-j)
\label{c20}
\\
\sigma^2_n & \leq & (c+5)\, u_2(0) 
\sum_{i} (\rho^2)^0_{ii} \, u_2(n-i) \ ,
\label{c30}
\end{eqnarray}
where $\sigma_n^2$ is defined in (\ref{c8})
and
$(\rho^2)^0_{ii}:=_{\,\, 0\!\!\!}\langle i | \rho^2|i\rangle_{\! 0}$.
Likewise, 
(\ref{b20}) and (\ref{b50}) yield
\begin{eqnarray}
& & \langle \rho_{mn}\rangle_V = 0
\label{c40}
\\
& & \langle |\rho_{mn}|^2 \rangle_V
=
\sum_{ik} |\rho_{ik}^0|^2 u_2(m-i) u_2(n-k)
\label{c50}
\end{eqnarray}
for all $m\not=n$.

Introducing (\ref{a101}) into (\ref{c20}) implies
\begin{equation}
\langle \rho_{nn}\rangle_V \leq
\max_j\{u_2(j)\} \sum_{j} \rho_{jj}^0 =u_2(0) \ ,
\label{c21}
\end{equation}
where we exploited that the sum over 
$j$ equals $\Tr \rho=1$.
Likewise, (\ref{c30}) yields
\begin{equation}
\sigma^2_n \leq  (c+5)\, u^2_2(0) \,\Tr\{\rho^2\} \ .
\label{c31}
\end{equation}
Rewriting the definition from (\ref{a135}) as
\begin{equation}
\Delta_n :=
\langle \rho_{n+1,n+1}\rangle_V-\langle \rho_{nn}\rangle_V
\label{c32}
\end{equation}
we find along similar lines of reasoning that
\begin{eqnarray}
|\Delta_n| 
& \leq & 
\sum_{j} \rho^0_{jj} |u_2(n+1-j)-u_2(n-j)|
\nonumber
\\
& \leq &
\max_n|u_2(n+1)-u_2(n)| \ .
\label{c33}
\end{eqnarray}
Moreover, we can conclude that
\begin{equation}
\sum_n |\Delta_n| 
\leq 
\sum_{j} \rho^0_{jj} \sum_n |u_2(n+1-j)-u_2(n-j)| \ .
\label{c34}
\end{equation}
Similarly as in (\ref{a140}), 
the last sum is seen to be equal to 
$2\,u_2(0)$. The remaining sum over $j$
equals $\Tr \rho=1$ and hence
\begin{equation}
\sum_n |\Delta_n| 
\leq 2\,u_2(0) \ .
\label{c35}
\end{equation}

Eqs. (\ref{c21}), (\ref{c31}) indicate that,
in contrast to pure states in (\ref{c1}),
for mixed states of small purity $\Tr\{\rho^2\}$,
the random fluctuations of the $\rho_{nn}$'s about
their mean values may become negligible.
Moreover, the right hand side of 
(\ref{c33}) usually turns out
\cite{deu91a,deu91b,fyo96,fla97,cas96}
to be of the order of $u_2^2(0)$.
The same conclusion is also suggested by (\ref{c35}).
Consequently, also the variations of 
$\langle\rho_{nn}\rangle_V$
as a function of $n$ become small.
Unlike for pure states we thus can now
conclude that (\ref{102}) and (\ref{101}) 
are approximately equal, implying thermalization.
However, assuming a small purity of $\rho$ 
represents a quite strong restriction in the 
first place.

Returning to general $\rho$, we can
deduce from (\ref{a101}) and (\ref{c20}) that
\begin{eqnarray}
\sum_n \langle \rho_{nn}\rangle_V^2 & \leq  &
\sum_n \max_j\{u_2(j)\}
\sum_{j} \rho_{jj}^0 \sum_{k} \rho_{kk}^0 u_2(n-k)
\nonumber
\\& \leq  &
u_2(0) \sum_{j} \rho_{jj}^0 \sum_{k} \rho_{kk}^0 
\sum_n u_2(n-k) \ .
\label{c60}
\end{eqnarray}
The sum over $n$ equals one according to
(\ref{a90}) and the two remaining sums are
equal to $\Tr \rho=1$, i.e.,
\begin{equation}
\sum_n \langle \rho_{nn}\rangle_V^2 
\leq  u_2(0) \ .
\label{c70}
\end{equation}
Likewise, Eqs. (\ref{c12}) and (\ref{c30}) imply
\begin{eqnarray}
\sum \sigma^2_n & \leq &
(c+5)\, u_2(0) 
\sum_{i} (\rho^2)^0_{ii} \, \sum_n u_2(n-i)
\nonumber
\\
& = & (c+5)\, u_2(0) \Tr\{\rho^2\} 
\leq (c+5)\, u_2(0) \ .
\label{c80}
\end{eqnarray}
Altogether, this yields
\begin{equation}
\sum_n \langle (\rho_{nn})^2\rangle_V \leq (c+6)\, u_2(0) 
\label{c90}
\end{equation}
and with (\ref{a115})
\begin{equation}
\langle \sum_n (\rho_{nn})^2\rangle_V \leq 10^{-\Or(N)} \ .
\label{c100}
\end{equation}
Since $(\max_n\rho_{nn})^2=\max_n(\rho_{nn})^2
\leq \sum_n (\rho_{nn})^2$ it follows that
\begin{equation}
\langle (\max_n \rho_{nn})^2\rangle_V \leq 10^{-\Or(N)} 
\label{c110}
\end{equation}
and hence by Markovs's inequality that
\begin{equation}
\mbox{Prob}\left(\max_n\{\rho_{nn}\}\geq\epsilon\right)\leq 
\epsilon^{-2}\, 10^{-\Or(N)} \ ,
\label{c120}
\end{equation}
see also the explanations below (\ref{a340}).

\subsection{Off-diagonals}
\label{s7c}
We focus on the off-diagonal matrix
elements $\rho_{mn}$ with $m\not=n$.
Their mean values are zero according 
to (\ref{c40}).
Their variance can be readily bounded by
the Cauchy-Schwarz inequality
\begin{equation}
|\rho_{mn}|^2\leq \rho_{mm}\, \rho_{nn} \ .
\label{c125}
\end{equation}
Likewise, introducing 
$|\rho^0_{ik}|^2\leq \rho_{ii}^0\rho_{kk}^0$ 
into (\ref{c50}) yields with (\ref{c20})
the estimate
\begin{equation}
\langle |\rho_{mn}|^2 \rangle_V 
\leq \langle \rho_{mm} \rangle_V \, \langle \rho_{nn} \rangle_V  \ .
\label{c130}
\end{equation}
Similarly as below (\ref{a285}) one sees 
that $\sum_{mn}|\rho_{mn}|^2=\Tr\{\rho^2\}$ 
and hence 
\begin{equation}
\sum_{m\not =n} \langle |\rho_{mn}|^2 \rangle_V
= \Tr\{\rho^2\} - \langle \sum_n\rho_{nn}^2\rangle_V 
\leq \Tr\{\rho^2\} \ .
\label{c139}
\end{equation}
According to (\ref{c100}), the last inequality
in (\ref{c139}) is
in fact a very tight upper bound.
The same estimate follows by summing in
(\ref{c50}) over $m$ and $n$.
Neither of these results indicate that the 
off-diagonal matrix elements are typically
much smaller than the diagonal elements.
We thus conjecture that typical off-diagonal
matrix elements will in fact not be small 
compared to the diagonal elements.
Trivial exceptions are pure states (\ref{c1}).
Non-trivial exceptions may be mixed 
states of low purity, similarly as 
below (\ref{c35}).

\subsection{Discussion}
\label{s7d}
The main result of this section is (\ref{c120}):
It implies that for the overwhelming
majority of randomly sampled perturbations $V$ 
in (\ref{a10}) the last term in (\ref{71}) is
unimaginably small (essentially in 
agreement with (\ref{12})).
In other words, equilibration in 
the sense of Sect. \ref{s2} is verified.
We emphasise that all these conclusions do not depend
an any further details of the actual initial 
condition $\rho(0)$, except that it is assumed 
to be fixed, i.e. independent of the randomly 
sampled $V$.

The physical interpretation is as follows:
One specific, usually not very well known, 
but nevertheless well-defined initial 
state $\rho(0)$ (pure or mixed) is ``given'' 
to us, and then evolves further 
in time according to one particular, 
randomly picked system Hamiltonian (\ref{a10}).
Our results guarantee that the vast
majority of those randomly sampled 
Hamiltonians gives rise to equilibration
in the sense of Eq. (\ref{71}).
As the only unproven part remains 
the assumption that the actual 
(in detail not exactly known) 
system does not correspond to one of
the rare, untypical Hamiltonians of 
the considered random ensemble.

Since $\Tr\{\rho\}=\sum_n\rho_{nn}=1$ 
we can conclude from (\ref{c120}) that,
for most $V$'s,
the number of non-negligible $\rho_{nn}$'s
{\em cannot} be much smaller than 
$10^{\Or(N)}$.
As a consequence, the strong ETH 
scenario from Sect. \ref{s5f} 
does not apply.
In turn, to apply this scenario, 
a different physical set up
is required, with a different
physical view of how the initial
condition $\rho$ arises.
Namely, one particular, but ``typical''
$V$ in (\ref{a10}) is considered 
to have been randomly sampled
but now is held fixed.
Since we assumed the system is 
typical, strong ETH as specified 
in Sect. \ref{s5f} can and 
will be taken for granted.
In a next step, the initial state 
$\rho=\rho(0)$ for this particular
system is specified, 
arising, e.g., as the result 
of an experimental preparation 
procedure for this very system
(with respect to other
systems $H$ of the ensemble, 
this preparation procedure may
not be physically meaningful or 
not even well defined).
Finally, there must be good reasons
(e.g. a very careful experimentalist)
to assume that this preparation
process yields level populations 
$\rho_{nn}$ which are negligible
outside a window of $n$-values 
much smaller than $10^{\Or(N)}$,
see Eq. (\ref{z100}).

In conclusion, our present 
formalism is able to validate
either equilibration or thermalization,
but not both of them simultaneously 
for one and the same physical model 
system.

\section{Comparison with the approach by Goldstein and coworkers}
\label{s8}
In a series of works \cite{gol10a,gol13,gol15,gol10b,gol11} 
Goldstein and coauthors
addressed the problem of thermalization 
by means of an approach which, at first
glance, seems to be entirely different from
our present one.
In particular, the ``intermediate'' problem of equilibration
(see Sect. \ref{s2}) apparently can be entirely 
circumvented.
One key point of their approach is the restriction
to so-called macroscopic observables, i.e.
observables whose statistical fluctuations are 
negligibly small for macroscopic systems 
at thermal equilibrium.
In this section, we will show that the latter 
defining property of a macroscopic observable 
implies that it satisfies (weak) ETH.
In other words, the restriction to such observables
is essentially equivalent to assuming ETH.

We define the microcanonical mean and variance 
of any given observable $A$ as
\begin{eqnarray}
\overline{A} & := & \Tr\{\rho_{\mic}\, A\}
\label{d10}
\\
\Delta A^2 & := & \Tr\{\rho_{\mic}\, (A-\overline{A})^2\} \ ,
\label{d20}
\end{eqnarray}
where $\rho_{\mic}$ is the microcanonical ensemble
from (\ref{z1}).
By definition, an observable $A$ is called a 
{\em macroscopic  observable}, if its fluctuations 
$\Delta A$ are negligibly small. 
A more precise formal requirement would be
vanishing fluctuations in the thermodynamic 
limit.
A more appropriate real world (experimentally
useful) version would be to require
that the fluctuations are
smaller than the experimental resolution 
limit $\dda$, with, e.g. $\dda=10^{-10}\da$, 
where $\da$ is the measurement range of 
the experimental instrument modelled by 
$A$, see Sect. \ref{s2}.

The above requirement represents a 
minimal condition:
Whatever alternative definition of a macroscopic 
observable may be proposed, if it admits non-small 
fluctuations in the microcanonical ensemble
then it would not seem well-defined to us.
Indeed, the definitions employed 
in \cite{gol10a,gol13,gol15} are
quite similar but not exactly identical to ours.
We also note that the microcanonical ensemble
itself  is only used here as a formal device to
define the notion of a macroscopic observable.
It does not in any way anticipate that the
actual system of interest should exhibit
thermalization.

Introducing (\ref{z1}) into (\ref{d10}) and (\ref{d20})
implies
\begin{eqnarray}
\overline{A} & = & \frac{1}{D}\sum_{\mic} A_{nn}
\label{d30}
\\
\Delta A^2 & = & \frac{1}{D}\sum_{\mic} (A^2)_{nn} - (\overline{A})^2 \ ,
\label{d40}
\end{eqnarray}
where $(A^2)_{nn}:=\langle n|A^2|n\rangle$.
In other words, (\ref{d30}) represents the average 
over those $A_{nn}$'s, 
whose energies $E_n$ are
contained in the microcanonical energy window.
Their typical deviation from this average is 
quantified by the variance
\begin{eqnarray}
\Delta A^2_{ETH} & := & \frac{1}{D}\sum_{\mic} (A_{nn} - \overline{A})^2
=\frac{1}{D}\sum_{\mic} (A_{nn})^2 - (\overline{A})^2 \ .
\label{d50}
\end{eqnarray}
From (\ref{a312}) we can conclude that 
$(A_{nn})^2\leq (A^2)_{nn}$ and hence that 
$\Delta A_{ETH} \leq \Delta A$.
Assuming that $A$ is a macroscopic observable
thus implies that $\Delta A_{ETH}$ is small
\cite{bri10}.
According to the definition in (\ref{d50})
it follows that most $A_{nn}$'s must be close 
to $\overline{A}$, i.e. weak ETH is satisfied
(see Sect. \ref{s5f}).

Closely related considerations are 
originally due to \cite{bri10}, 
however focusing on so-called intensive 
local few-body operators rather than on 
macroscopic observables.

Restricting oneself to macroscopic
observables clearly has a long and well 
founded tradition,
especially with respect to the thermodynamic 
roots of statistical physics.
On the other hand, statistical physics itself is
by no means restricted to such observables.
Rather, it is understood and experimentally 
(and numerically) seen that also 
``microscopic'' observables are
perfectly well described by this theory.
Such observables exhibit non-negligible 
fluctuations about their mean values and as 
such are hardly encountered in our everyday 
macroscopic world (exceptions may arise 
near critical points).
But already with the help of an optical
microscope, interesting observables 
exhibiting non-negligible thermal fluctuations 
(e.g. Brownian motion) become accessible.
Even more so, 
within the rapidly developing fields of nanotechnology 
and single molecule experiments, such microscopic
observables become of increasing practical relevance.

We finally remark that Goldstein
and coworkers conceive their own approach 
\cite{gol10a,gol13,gol15,gol10b,gol11} 
as a continuation of von Neumann's 
ground-breaking work \cite{neu29}.
Other relations between such an approach
and ETH than the one discussed above 
have also been
pointed out in Refs. \cite{gol11,rig12}.
Furthermore, the concept of
randomised Hamiltonians (or random matrices)
also plays a key role in the approach by 
von Neumann and by Goldstein and 
coworkers \cite{gol10b}.
However, rather than introducing 
this randomness directly into the 
Hamiltonian $H$ itself -- as done in 
the approach by Deutsch via Eq. (\ref{a10}) --
the randomness is now introduced
by prescribing the statistical properties 
of the randomly sampled eigenbases 
$|n\rangle$ of $H$, see 
Sect. \ref{s5b}, while keeping the
spectrum of $H$ fixed.
In spite of these technical 
differences, the two approaches 
are thus in fact very close in spirit
(see, e.g. Sect. 6 in \cite{gol10b}).

\section{Summary and conclusions}
\label{s9}
In a first step,
we reconsidered the random matrix model 
of Deutsch \cite{deu91a,deu91b,deu10}
and worked out a more detailed and slightly 
more general demonstration that it validates 
ETH:  For the overwhelming
majority of the corresponding random 
ensemble of Hamiltonians $H$, any given 
observable $A$ is represented in the
eigenbasis of $H$ by an almost 
diagonal matrix with very slowly varying 
diagonal elements.
More precisely: Apart from a fraction
of exceptional $H$'s which is exponentially 
small in the system size $N$,
the off-diagonal matrix elements  $A_{mn}$ 
are exponentially small in $N$ and the changes
of the diagonal elements $A_{nn}$ as a function
of $n$ are also exponentially small in $N$.
This implies the following solution
of problem (i) from Sect. \ref{s4}:
For any given $H$, one can readily
construct (a posteriori) an ETH-violating
$A$ (see Sect. \ref{s4}), but any such $A$ still 
continues to satisfy ETH for most other $H$'s.
In turn, if $H$ is not known in all details
with extremely high precision, then a 
given observable is exceedingly likely 
to exhibit ETH.

The generalisation for more than one 
observable is straightforward:
Given every single observable is exponentially 
unlikely to violate ETH, it is still extremely likely
that all of them will simultaneously 
exhibit ETH, as long as their number 
is not exponentially large,
i.e. remains within the limits of what
is feasible in any real (or numerical) 
experiment.

In a second step, we have shown by 
means of a further generalisation 
of Deutsch's approach
that also an essential prerequisite
for equilibration, namely Eq. (\ref{12}),
will be satisfied for the
overwhelming majority of 
Hamiltonians $H$.
In doing so, an arbitrary
(pure or mixed) state $\rho(0)$
is admitted as initial condition.
But this initial state $\rho(0)$ must
then remain always the same
for the entire ensemble
of random Hamiltonians $H$.

We also identified a not yet satisfactorily 
solved aspect of Deutsch's original approach 
and our present generalisation:
On the one hand, the changes of the
diagonal matrix elements $A_{nn}$ as a function
of $n$ are exponentially 
small in the system size $N$
up to exponentially rare exceptions.
On the other hand, the typical
difference between neighbouring 
energy levels $E_n$ is also exponentially
small in $N$ (cf. Sect. \ref{s2}),
i.e. the number of energy eigenvalues 
contained in 
an energy intervals $[E-\Delta E,\, E]$ 
is exponentially large in $N$
for the usual $\Delta E$'s of interest.
Hence, the variations
of $A_{nn}$ within the entire energy
interval may no longer be negligible.
As a consequence, thermalization, i.e.,
the practical indistinguishability of (\ref{102})
and (\ref{101}), can only be proven
under the extra condition that
the interval of relevant $n$-values, 
which notably contribute to those 
sums in (\ref{102}) and (\ref{101}), 
is not too large, namely much 
smaller than $10^{\Or(N)}$
(cf. (\ref{z100})).
In other words, only exceedingly 
small $\Delta E$'s are admitted.
In the following four paragraphs, 
we conclude with four noteworthy
remarks and implications.

In spite of this restriction, the admitted
range of $n$-values is still huge,
and likewise for the admitted energy 
intervals $\Delta E$ in comparison
with the energy level spacings.
In particular, they are still of physical 
interest:
For instance, one may imagine that the
experimentalist has prepared the system
with a sufficiently small uncertainty in the
total system energy $E$ so that
the corresponding condition can be
safely taken for granted for the 
$\rho_{nn}$'s appearing in 
(\ref{102}) and (\ref{101}).

As mentioned already in in Sect. \ref{s4},
our findings imply the equivalence of 
$\rho_{eq}$ in (\ref{101}) not only with
the microcanonical ensemble 
$\rho_\mic$ in (\ref{102}) but also
with any other equilibrium (i.e. steady state)
ensemble, provided that its level populations 
are mainly concentrated within a sufficiently
small energy window as specified above.
Unfortunately, this condition is not
satisfied e.g. for the canonical 
ensemble.

The other, more fortunate side of 
the coin is that within our present 
approach the diagonal matrix 
elements $A_{nn}$ are indeed
{\em not} forbidden to exhibit 
non-negligible variations
for sufficiently large changes 
of $n$, or, equivalently,
for macroscopically 
notable changes of $E_n$.
This solves problem (ii) from Sect. \ref{s4}.

Assuming one and the same initial state 
$\rho(0)$ for the entire ensemble of 
random Hamiltonians $H$, as done
in our above discussion of 
equilibration, implies that
the number of $\rho_{nn}$'s
which notably contribute 
in (\ref{101}) is {\em not} much smaller 
than $10^{\Or(N)}$ for most $H$'s.
In conclusion, our
present generalisation
of Deutsch's approach allows us to 
corroborate either equilibration or 
thermalization, but not both of them 
simultaneously for one and the same 
physical model system. 
The root of the problem is as before:
Whether and why the dependence of the
diagonal matrix elements $A_{nn}$ on 
$n$ is neither too strong nor too weak
is not yet fully satisfactorily understood.
A solution of this problem is presently 
being worked out.

\section*{Acknowledgments}
This work was supported by DFG-Grant RE1344/7-1.
We acknowledge support for the Article 
Processing Charge by the Deutsche Forschungsgemeinschaft
and the Open Access Publication Fund of Bielefeld University.

\end{document}